\documentclass{article}
\usepackage{amssymb}

\begin{document}

\title{Dark Matter from a gas of wormholes}
\author{A. A. Kirillov, E.P. Savelova \\
%EndAName
\emph{Branch of Uljanovsk State University in Dimitrovgrad, }\\
\emph{Dimitrova str 4.,} \emph{Dimitrovgrad, 433507, Russia} }
\date{}
\maketitle

\begin{abstract}
The simplistic model of the classical spacetime foam is considered, which
consists of static wormholes embedded in Minkowski spacetime. We explicitly
demonstrate that such a foam structure leads to a topological bias of
point-like sources which can equally be interpreted as the presence of a
dark halo around any point source. It is shown that a non-trivial halo
appears on scales where the topological structure possesses  local
inhomogeneity, while the homogeneous structure reduces to a constant
renormalization of the intensity of sources. We also show that in general
dark halos possess both (positive and negative) signs depending on scales
and specific properties of the topological structure of space.
\end{abstract}

\section{Introduction}

The discrepancy between the luminous matter and the dynamic, or gravitating
mass was first identified in clusters of galaxies \cite{Zw}. Since then it
has widely been accepted that the leading contribution to the matter density
of the Universe comes from a specific non-baryonic form of matter (Dark
Matter (DM), e.g., see Refs. \cite{dm}). Apart from some phenomenological
properties of DM ( it starts to show up in galactic halos, it is
non-baryonic, it is cold at the moment of recombination, it remains to be
cold in clusters and at larger scales (e.g., see Ref. \cite{Pr} and
references therein), but in a strange way, it turns out to be worm in
galaxies\footnote{%
Cold DM should necessary \ form a cusp in centers of galaxies $\rho
_{DM}\sim 1/r$ \cite{NFW}, while observations definitely show the cored
distribution $\rho _{DM}\sim const$ \cite{Core}. The only way to destroy the
cusp and get the cored distribution is to introduce some self-interaction in
DM or to consider worm DM. Both possibilities are rejected at large scales
by observing $\Delta T/T$ spectrum (e.g., see Ref. \cite{Pr} and references
therein).} \cite{Core}, etc.) nothing is known of its nature. Particle
physics suggests various hypothetical candidates for Dark Matter, while we
still do not observe such particles in direct laboratory experiments.
Moreover, DM displays so non-trivial properties (it is worm or
self-interacting in galaxies, however it was cold at the moment of
recombination and it is still cold on larger (than galaxies) scales, DM
fraction is practically absent in intracluster gas \cite{DMpaper}), etc.,
that it is difficult to find particles capable of reconciling such
observations. These facts suggest us to try, as an alternative to DM
hypothesis, the possibility to interpret the observed discrepancy between
luminous and gravitational masses as a violation of the law of gravity.

Possible violations of the gravity law (or modifications of general
relativity (GR) have widely been discussed, e.g., see Refs. \cite{VN,sand}.
The common feature of such theories is the presence of some characteristic
energy scale $E_{0}$ (e.g., some kind of a mass of gravitons Refs. \cite%
{sand} or even a fundamental acceleration in the modified Newtonian dynamics
\cite{MOND}) which represents the threshold upon which DM - type phenomena
(violations of the gravity law) start to show up. However, DM features
pointed out above clearly indicate a non-linearity of DM phenomena and that
there cannot be a single fundamental scale in DM physics. Again,
observations demonstrate that DM halos have different properties
(distributions) in different galaxies which also cannot be prescribed to a
single fundamental scale. In other words, it turns out to be rather
difficult to get a modification of GR which is flexible enough to reconcile
all the observational DM data. Moreover, there exist fundamental theoretical
arguments (e.g., the massless nature of gravitons, etc.) which make any
modification to be undesirable from particle physics standpoint.

It turns out however, that a modification of the theory itself is not the
only possibility to violate the Newton's law. The standard Newton's law can
easily be modified when the topological structure of space is different from
$R^{3}$. In the first place the nontrivial topological structure was shown
to display itself by the topological bias of all physical sources (e.g., see
Ref. \cite{K06} and references therein) which is equivalent to the presence
of DM. Moreover, there exist the very basic theoretical arguments in favour
of the presence of a non-trivial topological structure of space. Indeed, as
it was first suggested by Wheeler, at Planck scales spacetime should undergo
quantum topology fluctuations (the so-called spacetime foam) \cite{wheeler}.
Such fluctuations were strong enough to form a foam-like structure of space
during the quantum stage of the evolution of the Universe. There are no
convincing theoretical arguments of why such a foam-like structure should
decay upon the quantum stage. Moreover, the presence of a considerable
portion of Dark Energy in the present Universe \cite{Pr} (and in the past,
on the inflationary stage) may serve as the very basic indication of a
nontrivial topological structure of space\footnote{%
Recall that DE violates the energy domination condition. Save speculative
theories (or pure phenomenological models), there is no matter which meets
such a property. However in the presence of a non-trivial topology, vacuum
polarization effects are known to give rise quite naturally to such a form
of matter. By other words, up to date the only rigorous way to introduce
Dark Energy is to consider the vacuum polarization effects on manifolds of a
non-trivial topological structure.}.

We note that inflationary stage in the past \cite{inf} should enormously
stretch all physical scales and, therefore, we should expect relics of the
primordial foam-like structure to survive at very large (astronomically
considerable) scales. The foam-like structure, in turn, was shown to be
flexible enough to account for the all the variety of DM phenomena (e.g.,
see Refs. \cite{K06,KT07}), for parameters of the foam may arbitrary vary in
space to produce the observed variety of DM halos in galaxies (e.g., the
universal rotation curve for spirals constructed in Ref. \cite{KT06} on the
basis of the topological bias perfectly fits observations). Moreover, the
topological nature of the bias means that the DM halos surrounding
point-like sources appear due to the scattering on topological defects (on
the foam-like structure) and if a source radiates, such a halo turns out to
be luminous too \cite{KT07} which seems to be the only way to explain
naturally the observed absence of DM fraction in intracluster gas \cite%
{DMpaper}.

A general foamed Universe can be viewed as the standard Friedman model
filled with a gas of wormholes \cite{KT07}. However, a priori it is not
clear if the presence of such a gas is suffficient to get DM phenomena. In
the present paper we consider the simplistic exact model of the spacetime
foam, which consists of a static gas of wormholes embedded in the Minkowski
space and demonstrate how basic DM effects can be explicitly evaluated. We
note that simplistic models of the spacetime foam have been already
considered in the literature (e.g., see Ref. \cite{BK07} and references
therein where also other topological defects were considered). However the
primary interest was there focused on setting observational bounds on the
foam-like structure at extremely small scales (which correspond to the
energies higher than $100$ $Mev$), while DM phenomena suggest that the
characteristic scale of the spacetime foam (and respectively of wormholes)
should be of the galaxy scale, e.g., of the order of a few $Kpc$. The
rigorous bounds obtained indicate that at small scales spacetime is
extremely smooth up to the scales $\gtrsim 10^{2}L_{pl}$ (where $L_{pl}$ is
the Planck length), that was to be expected. Indeed, at those scales
topology fluctuations have only virtual character and due to the
renormalizability of physical field theories should not directly contribute
to observable (already renormalized) effects. Topology fluctuations were
strong only during the quantum stage of the evolution of the Universe, while
the subsequent inflationary phase considerably increases all characteristic
scales of the foam. Therefore, the only possibility to find effects of the
relic foam - like structure of space is to seek for them at very large
scales, rather than at very small ones.

\section{Modification of the Newton's law in the presence of a single
wormhole}

In the present section we, for the sake of simplicity, consider the flat $%
R^{3}$ space. We consider first a single wormhole, which represents a couple
of conjugated spheres $S_{\pm }$ of the radius $a$ and with a distance $%
d=\left\vert \vec{R}_{+}-\vec{R}_{-}\right\vert $ between centers of
spheres. The interior of the spheres is removed and surfaces are glued
together. Our aim is to find the Green function $\triangle G(r,r_{0})=4\pi
\delta (r-r_{0})$ for such a topology.

In the absence of the wormhole the solution is well known $G_{0}(r)=-1/r$
which represents the standard Newton's law. In the case of a non-trivial
topology of space (i.e., in the presence of the wormhole) the Newton's law
violates, however, we still can use the standard Green function (the
standard Newton's law), while the nontrivial topology (i.e. the proper
boundary conditions) will be accounted for by the topological bias of the
source \cite{K06,KT07} $\delta (r-r_{0})$ $\rightarrow $ $\delta (r-r_{0})$~$%
+$ $b\left( r,r_{0}\right) $, where $b\left( r,r_{0}\right) $ $=$ $\sum
e_{A}\delta (r-f_{A}(r_{0}))$ describes ghost images which produce the
topological corrections to the Newton's law. The equivalent description is
the introduction of the topological permeability $\widehat{\varepsilon }$,
i.e. the modification of the equation itself $\triangle \widehat{\varepsilon
}G(r,r_{0})=4\pi \delta (r-r_{0})$ which gives\footnote{%
In the case of homogeneous and isotropic topological structure (i.e., when $%
b=b\left( \left\vert r-r_{0}\right\vert \right) $) the relation between $b$
and $\widehat{\varepsilon }$ is trivial in the Fourier representation, i.e. $%
\varepsilon (k)=1/\left( 1+b\left( k\right) \right) $.} $G(r,r_{0})$ $=$ $-%
\widehat{\varepsilon }^{-1}(1/\left\vert r-r_{0}\right\vert )$ $=$ $%
-1/\left\vert r-r_{0}\right\vert $ $-\int b\left( r,r^{\prime }\right)
/\left\vert r^{\prime }-r_{0}\right\vert dV^{\prime }$. In the situation
when $\widehat{\varepsilon }=const$ (e.g., at very large scales) the
topological permeability renormalizes merely the value of a source $%
\triangle G(r,r_{0})=(4\pi /\widehat{\varepsilon })\delta (r-r_{0})$ (or
equivalently the value of the interaction constant $\gamma \rightarrow
\gamma /\widehat{\varepsilon }$ \cite{K03}).

We note that the wormhole can be equally viewed as a couple of spherical
conjugated mirrors, so that while the incident signal falls on one mirror
the reflected signal comes from the conjugated mirror. We point also out
that gas of spherical mirrors has many common features with the gas of
wormholes. In particular, in the case of a homogeneous distribution of
stationary sources, statistical properties of the Newton or gravitational
potential remain the same. As we shall see the difference appears only when
we consider the topological permeability of space. In the case of wormholes
we can distinguish two types of prmeabilities\footnote{%
As it will be shown latter, in the case of mirrors the susceptibility of
space $\chi $ $=$ $(\widehat{\varepsilon }-1)/4\pi $ is always negative
(anti-screening $\chi <0$), i.e., the polarizability is opposite to the
external field , while for wormholes there appear two types of polarization (%
$\chi >0$ and $\chi <0$). In analogy with the magnetic susceptibility one
can speak of dia- and para- susceptibilities of space.}: one which gives $%
\widehat{\varepsilon }<1$ and the second part with $\widehat{\varepsilon }>1$%
, while mirrors possess only one type of permeability ($\widehat{\varepsilon
}<1$, i.e., anti-screening). The difference appears also when we consider
the propagation of signals in such a medium, which we will discuss
elsewhere. However one spherical mirror gives the simplest example of a
non-trivial topology and from the methodical standpoint it is more
convenient to start with this case\footnote{%
The simplest example is given by a plane mirror, but this case is trivial
and we think that every reader can reconstruct such a case by himself.}
which we latter on generalize to the case of a wormhole.

\subsection{The case of a single spherical mirror}

Consider a spherical mirror of the radius $a$ and at the position $\vec{R}$.
Then points $|\vec{r}-\vec{R}|<a$ represent the non-physical region and the
exact form of the topological bias and the Green function depends on the way
of how we continue coordinates in the non-physical region. We need not to
say that the values of the Green function in the physically admissible
region $|\vec{r}-\vec{R}|\geq a$ do not depend on the continuation at all.
However the bias and the form of the Green function in the "non-physical"
region do depend on this.

First, we consider the continuation which we use in astrophysics. Recall
that in astrophysics we map the physical space $\mathcal{M}$ onto $R^{3}$ as
follows \cite{KT07}. We take a point $O$ in space (the position of an
observer) and issue geodesics from $O$ in every direction. Then points in $%
\mathcal{M}$ can be labeled by the distance from $O$ and by the direction of
the corresponding geodesic. In other words, for an observer at $O$ the space
will always look as $R^{3}$. However if we take a point $P\in \mathcal{M}$,
there may exist many homotopically non-equivalent geodesics connecting $O$
and $P$. Thus, the point $P$ will have a number of images in $R^{3}$. Recall
that the observer might determine the topology of $\mathcal{M}$ by noticing
that in the observed space $R^{3}$ there is a fundamental domain $\mathcal{D}
$ such that every radiation or gravity source in $\mathcal{D}$ has a number
of copies $N$ outside $\mathcal{D}$. Then the actual manifold $\mathcal{M}$
is obtained by identifying the copies $R^{3}/N$.

In the case of one spherical mirror every point in the physically admissible
region has only one copy in the non-physical region $|\vec{r}-\vec{R}|<a$
which corresponds to the one-to-one map (i.e., the reflection law) $\vec{r}$
$\rightarrow $ $\vec{f}(r)$ $=$ $\vec{R}$ $+$ $a^{2}(\vec{r}-\vec{R})/(\vec{r%
}-\vec{R})^{2}$. In this picture the interior of the sphere $|\vec{r}-\vec{R}%
|\leq a$ is absolutely equivalent to the outer region, i.e., $|\vec{f}\left(
\vec{r}\right) -\vec{R}|\geq a$ and the metric within the sphere $|\vec{r}-%
\vec{R}|\leq a$ is flat and is given by $dl^{2}=d\vec{f}^{2}\left( r\right) $%
. In particular, in this case the volume within the sphere is infinite, for
it coincides exactly with the volume of the outer region of the sphere.
However while in the outer region geodesics are straight lines in the inner
region geodesics are represented by circles which go through the center of
the sphere. In such a picture every source of gravity at the position $\vec{r%
}_{0}$ ($r_{0}>a$) will be accompanied with the only source at the position $%
\vec{r}_{1}=\vec{f}(r_{0})$ within the sphere ($\left\vert
r_{1}-R\right\vert <a$).

In the present paper we however will use the more standard way when the
sphere represents merely a portion of $R^{3}$ with the standard flat metric (%
$dl^{2}=d\vec{r}^{2}$) within it and the volume of the sphere being $V(a)=%
\frac{4}{3}\pi a^{3}$. In such a case we can use the inversion method (see
the standard books, e.g., Ref. \cite{jackson}). In the case of one spherical
mirror the proper boundary conditions can be satisfied if we place within
the sphere a couple of odd image ("ghost") sources, i.e.,%
\begin{equation}
\delta (\vec{r}-\vec{r}_{0})\rightarrow \delta (\vec{r}-\vec{r}_{0})+\frac{a%
}{y}\delta (\vec{r}-\vec{r}_{1})-\frac{a}{y}\delta (\vec{r}-\vec{R}),
\label{b1}
\end{equation}%
where $\vec{r}_{1}=\vec{R}+\frac{a^{2}}{y^{2}}\vec{y}$ and $\vec{y}=\vec{r}%
_{0}-\vec{R}$. The negative source, at the center of the sphere, is here
added to compensate the reflected source at $\vec{r}_{1}$. Physically, this
means that the mirror does not radiate (virtual photons or gravitons) itself
but only redistributes the existing radiation. In the electrodynamics this
means that such a medium (gas of mirrors) possesses some polarization
property which gives rise to the origin of magnetic and dielectric
permeabilities Ref. \cite{jackson} (see also Refs. \cite{BK07}).

Thus (\ref{b1}) defines the topological bias in the form%
\begin{equation}
b\left( r,r_{0}\right) =b^{\left( +\right) }-b^{\left( -\right) }=\frac{a}{y}%
\left( \delta (\vec{r}-\vec{r}_{1})-\delta (\vec{r}-\vec{R})\right)
\label{b2}
\end{equation}%
which has the property $\int b\left( r,r_{0}\right) d^{3}r=0$. We see that
the bias is solely defined in the non-physical region (interior of the
sphere) and therefore its values depend essentially on the way of
continuation discussed. When we use the astrophysical way the interior and
the outer region of the sphere are simply coincide and we need not to
introduce the additional negative source\footnote{%
On the surface of the mirror the negative surface source is automatically
generated, for the distance between opposite points on the sphere is
infinite.} and we will get $\int_{\left\vert r-R\right\vert <a}b\left(
r,r_{0}\right) d^{3}r$ $\equiv $ $\int_{\left\vert r-R\right\vert >a}\delta (%
\vec{r}-\vec{r}_{0})d^{3}r$ $=$ $1$.

Thus, in the physically admissible region ($|\vec{r}-\vec{R}|\geq a$) the
exact form of the Green function is given by%
\begin{equation}
-G(r)=\frac{1}{\left\vert \vec{r}-\vec{r}_{0}\right\vert }+\frac{a}{y}\frac{1%
}{\left\vert \vec{r}-\vec{r}_{1}\right\vert }-\frac{a}{y}\frac{1}{\left\vert
\vec{r}-\vec{R}\right\vert },  \label{g1}
\end{equation}%
while its form in the non-physical region ($|\vec{r}-\vec{R}|<a$) depends
essentially on the continuation procedure. The standard continuation gives
the same expression (\ref{g1}) for the non-physical region $|\vec{r}-\vec{R}%
|<a$, while the continuation by the astrophysical way gives $G\rightarrow
G\left( f\left( r\right) \right) $ (i.e., while $r$ runs the nonphysical
region $|\vec{r}-\vec{R}|<a$, $f\left( r\right) $ runs the region $|\vec{f}%
\left( r\right) -\vec{R}|>a$).

\subsection{The case of a wormhole}

In the case of a wormhole we have a couple of conjugated mirrors, so that
while the incident signal falls on one mirror the reflected signal comes
from the conjugated mirror. Thus, we have to replace the positive image
source in (\ref{b1}) into the conjugated sphere and rotate it with some
matrix $U$, which defines the gluing procedure. Moreover, every image\ (
ghost sources, positive and negative ones) undergoes again the reflections
upon the conjugated mirror and thus produces a countable set of images. Let $%
\vec{R}_{+}$, $\vec{R}_{-}$ be the vectors for the positions of centers of
the spheres and let us define the transformations%
\begin{equation}
\ \vec{r}_{\pm 1}=T_{\pm }\vec{r}_{0}=\vec{R}_{\pm }+\frac{a^{2}}{(\vec{r}%
_{0}-\vec{R}_{\mp })^{2}}U^{\pm 1}(\vec{r}_{0}-\vec{R}_{\mp }).  \label{r+}
\end{equation}%
Applying such a transformation many times we get for the positions of extra
positive images%
\begin{equation}
\ \vec{r}_{\pm n}=T_{\pm }^{n}\vec{r}_{0}=\vec{R}_{\pm }+\frac{a^{2}}{(\
T_{\pm }^{n-1}\vec{r}_{0}-\vec{R}_{\mp })^{2}}U^{\pm 1}(T_{\pm }^{n-1}\vec{r}%
_{0}-\vec{R}_{\mp }),
\end{equation}%
which define the positive part of the topological bias in the form
\begin{equation}
b^{\left( +\right) }\left( r,r_{0}\right) =\sum_{\ n=0}^{\infty
}b_{+n}^{\left( +\right) }\delta (\vec{r}-T_{+}^{n+1}\ \vec{r}%
_{0})+b_{-n}^{\left( +\right) }\delta (\vec{r}-T_{-}^{n+1}\ \vec{r}_{0}),
\label{b}
\end{equation}%
where%
\begin{equation}
b_{\pm n}^{\left( +\right) }=\prod\limits_{m=0}^{n}\frac{a}{\left\vert
T_{\pm }^{m}\vec{r}_{0}-\vec{R}_{\mp }\right\vert }.  \label{pr}
\end{equation}%
In the analogous way by means of the use of the transformation (\ref{r+})
and starting with sources $\frac{a}{\left\vert \vec{r}_{0}-\vec{R}_{\pm
}\right\vert }\delta (\vec{r}-\vec{R}_{\pm })$ we define the negative part
of the bias $b^{\left( -\right) }\left( r,r_{0}\right) $. We note that all
images $\vec{r}_{\pm n}$ lie within the respective spheres $S_{\pm }$.

The above expressions solve the problem posed and the exact Green function
(e.g., the gravitational potential for a point source at the position $r_{0}$%
) is given by
\[
-G(r)=1/\left\vert r-r_{0}\right\vert +\sum b_{\pm n}^{\left( +\right)
}/\left\vert r-r_{\pm n}\right\vert -\sum b_{\pm m}^{\left( -\right)
}/\left\vert r-r_{\pm m}^{\left( -\right) }\right\vert .
\]%
Here the first term represents the standard Newton's law, while the sums
describe topological corrections, which in observations can be equally
interpreted as the presence of some amount of extra (or dark) matter.

Consider now the degree of polarization of space produced by a wormhole.
Since by definition the topological bias has the property $\int
b(r,r_{0})d^{3}r\equiv 0$ and the characteristic distance between spheres is
$d=\left\vert \vec{R}_{+}-\vec{R}_{-}\right\vert $ it can be expressed by
the positive part of the bias, i.e.,
\begin{equation}
Q^{\left( +\right) }=\int b^{\left( +\right) }\left( r,r_{0}\right)
d^{3}r=\sum_{\ n=0}^{\infty }b_{+n}^{\left( +\right) }+b_{-n}^{\left(
+\right) }.
\end{equation}%
Consider the first term of this sum, i.e., $\sum b_{+n}^{\left( +\right)
}=I_{+}$. It is convenient to extract the common multiplier $I_{+}$ $=$ $%
b_{+0}^{\left( +\right) }$ $(1+\sum_{n=1}^{\infty }b_{+n}^{\left( +\right)
}/b_{+0}^{\left( +\right) })$, where $b_{+0}^{\left( +\right) }=a/\left\vert
\vec{r}_{0}-\vec{R}_{-}\right\vert $ depends essentially on the positions of
the source and the wormhole. Suppose that the wormhole obeys the condition $%
d\gg a$. Then in the product (\ref{pr}) for $m\geq 1$ we can use the
approximation $\left\vert T_{\pm }^{m}\vec{r}_{0}-\vec{R}_{\mp }\right\vert
\approx \left\vert \vec{R}_{\pm }-\vec{R}_{\mp }\right\vert =d$ (the next
terms have the order $a/d\ll 1$). In this approximation coefficients (\ref%
{pr}) take the form%
\begin{equation}
b_{\pm n}^{\left( +\right) }\simeq b_{\pm 0}^{\left( +\right) }\left( \frac{a%
}{d}\right) ^{n}  \label{bn}
\end{equation}%
and the sum gives%
\begin{equation}
Q^{\left( +\right) }=\left( b_{+0}^{\left( +\right) }+b_{-0}^{\left(
+\right) }\right) \sum_{\ n=0}^{\infty }\left( \frac{a}{d}\right)
^{n}=\left( \frac{a}{\left\vert \vec{r}_{0}-\vec{R}_{-}\right\vert }+\frac{a%
}{\left\vert \vec{r}_{0}-\vec{R}_{+}\right\vert }\right) \frac{d}{d-a}.
\label{q}
\end{equation}

The factor $d/(d-a)\approx 1$ describes corrections of multiple reflections
of images while the leading contribution comes from the first order images.
We recall that by the construction the source lies always outside the
spheres, which means that $a/\left\vert \vec{r}_{0}-\vec{R}_{\pm
}\right\vert \leq 1$ (the equality can be achieved only when the source
comes close to one of the spheres $S_{\pm }$). Thus we see that the
amplitude of additional sources produced by a single wormhole may reach the
order $\sim 1$.

The expression (\ref{bn}) shows that the intensity of a ghost source which
corresponds to the multiple reflection (of the order $n$) decreases as $%
\left( a/d\right) ^{n}$ (recall that $a/d<1$). Then for wormholes obeying
the condition $a/d\ll 1$ it is sufficient to retain only the first order
images which define the bias
\begin{equation}
b\left( r\right) =\frac{a}{R_{-}}\left[ \delta (\vec{r}-\vec{r}_{+1})-\delta
(\vec{r}-\vec{R}_{-})\right] +\frac{a}{R_{+}}\left[ \delta (\vec{r}-\vec{r}%
_{-1})-\delta (\vec{r}-\vec{R}_{+})\right]  \label{bk}
\end{equation}%
where we have used the coordinate system in which $r_{0}=0$ and values $\vec{%
r}_{\pm 1}$ are defined by (\ref{r+}).

\section{Static gas of wormholes}

In what follows we, for the sake of simplicity, will assume that the source
is at the origin $r_{0}=0$. First, we consider some general qualitative
properties of the bias which can be obtained from simple geometric
consideration and which should be valid also in more general situations
(e.g., when the non-trivial topology cannot be reduced to a gas of
wormholes).

Indeed, the basic effect of a non-trivial topology is that it cuts some
portion of the volume of the coordinate space. Therefore, the volume of the
physically admissible region becomes smaller, while the density of virtual
gravitons/photons (or equivalently, the density of lines of the strength of
force) becomes higher. From the standard flat space standpoint this will
effectively look as if the amplitude of a source renormalizes. Let $M$ be
the value of the source and consider a ball of the radius $r$ around the
source. Then the physical volume of the ball is $V_{ph}\left( r\right) $ $=$
$V_{coor}\left( r\right) $ $-$ $V_{m}\left( r\right) $, where the coordinate
volume is $V_{coor}$ $=$ $\left( 4\pi /3\right) r^{3}$ and $V_{m}\left(
r\right) $ is the volume of mirrors or wormholes which get into the ball.
Therefore, the actual value of the surface which restricts the ball is given
by $S_{ph}\left( r\right) $ $=$ $\frac{d}{dr}V_{ph}\left( r\right) $. Then
we can use the Gauss divergency theorem to estimate the renormalization of
the source. Indeed, the Gauss theorem states that%
\[
\int\limits_{S\left( R\right) }n\nabla GdS=4\pi \int_{r<R}M\delta \left(
r\right) dV=4\pi M,
\]%
where $G$ is the true Green function. Then for an isotropic distribution of
wormholes it defines the normal projection of the force as $F_{n}\left(
R\right) =n\nabla G=4\pi M/S_{ph}\left( R\right) $.

This can be rewritten as in the ordinary flat space (i.e., in terms of the
standard Green function $G_{0}=-1/r$ and the coordinate surface $%
S_{coor}=4\pi R^{2}$) $F_{n}\left( R\right) =M^{\prime }\left( R\right)
/R^{2}$, where $M^{\prime }\left( R\right) =4\pi R^{2}M/S_{ph}\left(
R\right) $ which defines the bias in the form $M^{\prime }\left( R\right)
/M=1+4\pi \int_{0}^{R}b\left( r\right) r^{2}dr$ or%
\begin{equation}
b\left( r\right) =\frac{1}{r^{2}}\frac{d}{dr}\frac{r^{2}}{\frac{d}{dr}%
V_{ph}\left( r\right) }.  \label{BT}
\end{equation}

Thus, we see that a non-trivial bias $b\left( r\right) $ appears, in the
fist place, due to the discrepancy in the behavior of the physical volume $%
V_{ph}\left( r\right) $ and that of $V_{coor}\left( r\right) $. At scales
where the distribution of wormholes (or mirrors) crosses over to homogeneity
we get $\overline{V}_{ph}\left( R\right) $ $=$ $\varepsilon V_{coor}\left(
R\right) $ $=4/3\pi R^{3}\varepsilon $ with a constant $\varepsilon <1$.
This gives $\overline{b}\left( r\right) =0$ at such scales, but defines the
renormalization of the point source as $M^{\prime }/M$ $=$ $1/\varepsilon $.

Consider now a set of wormholes with parameters $\vec{R}_{n,\pm }$, $U_{n}$
and $a_{n}$ ($n=1,2,...N$). We shall assume that the gas is sufficiently
rarefied (i.e., $n\ll 1$ where $n=N/V$ is the density of wormholes).
Therefore, we can neglect the feedback of wormholes, i.e., images which
appear due to the reflection between wormholes, and evaluate the
permeability of space $\varepsilon $ (and the bias $b\left( r\right) $ ) in
the linear approximation for the external field of the form $\phi =-1/r$.
The permeability of a dense gas can then be obtained in the standard way.
Indeed, if we present $\varepsilon =1+4\pi \chi $, where $\chi $ is the
susceptibility of space, then for a dense gas it is related to the linear
susceptibility $\chi _{0}$ as $\chi =\chi _{0}/\left( 1-4/3\pi \chi
_{0}\right) $, e.g., see part 4. in Ref. \cite{jackson}.

It is convenient to distinguish in (\ref{bk}) the two parts of the bias $%
b=b_{0}+b_{1}$, where $b_{0}$ resembles the bias of the spherical mirrors (%
\ref{b2})
\begin{equation}
b_{0}\left( r\right) =\sum_{\sigma =\pm }\frac{a}{R_{\sigma }}\left[ \delta (%
\vec{r}-\vec{r}_{\sigma 1})-\delta (\vec{r}-\vec{R}_{\sigma })\right] ,
\label{bk0}
\end{equation}%
while the rest part is given by
\begin{equation}
b_{1}\left( r\right) =a\left( \frac{1}{R_{+}}-\frac{1}{R_{-}}\right) \left[
\delta (\vec{r}-\vec{r}_{-1})-\delta (\vec{r}-\vec{r}_{+1})\right] .
\label{bk1}
\end{equation}%
Both parts give different contributions to $\varepsilon $ and should be
considered separately.

\subsection{The case $\protect\varepsilon <1$}

Consider first the part of the bias (\ref{bk0}) which coincides formally
with the bias produced by a gas of spherical mirrors. In this case the
topological bias is%
\begin{equation}
b_{0}\left( r\right) =\sum \frac{a_{n}}{R_{\pm ,n}}\left[ \delta (\vec{r}-%
\vec{r}_{\pm ,n})-\delta (\vec{r}-\vec{R}_{\pm ,n})\right]   \label{bk0n}
\end{equation}%
where $r_{\pm n}^{\alpha }=$ $R_{\pm ,n}^{\alpha }-a_{n}^{2}/R_{\mp
,n}^{2}U_{n,\alpha \beta }^{\pm 1}R_{\mp ,n}^{\beta }$, see (\ref{r+}). We
shall assume that for all wormholes $a/R_{\pm ,n}\ll 1$ and, therefore, (\ref%
{bk0n}) can be expanded by the small parameter $a/R_{\pm }$ which gives
\begin{equation}
b_{0}\left( r\right) =\nabla _{\alpha }\sum U_{n,\alpha \beta }^{\pm
1}R_{\mp ,n}^{\beta }\frac{a^{3}}{R_{\pm ,n}R_{\mp ,n}^{2}}\delta (\vec{r}-%
\vec{R}_{\pm ,n})  \label{p}
\end{equation}%
where $\nabla _{\alpha }=\partial /\partial r^{\alpha }$.

Let $F\left( R_{\pm },a,U\right) $ be the density of wormholes with
parameters $R_{-}$, $R_{+}$, $U$ and $a$, i.e.,
\begin{equation}
F\left( R_{\pm },a,U\right) =\sum\limits_{n}\delta \left( \vec{R}_{-}-\vec{R}%
_{-}^{n}\right) \delta \left( \vec{R}_{+}-\vec{R}_{+}^{n}\right) \delta
\left( a-a_{n}\right) \delta \left( U-U_{n}\right)   \label{F}
\end{equation}%
which allows to rewrite (\ref{p}) in the form%
\begin{equation}
b_{0}\left( r\right) =\nabla _{\alpha }\frac{1}{r}\sum_{s=\pm }\int \frac{%
R_{-s}^{\beta }}{R_{-s}^{2}}\delta (\vec{r}-\vec{R}_{s})H_{\alpha \beta
}^{s}\left( \vec{R}_{+},\vec{R}_{-}\right) d^{3}R_{+}d^{3}R_{-},  \label{H}
\end{equation}%
where%
\[
H_{\alpha \beta }^{\pm }\left( R_{+,}R_{-}\right) =\int a^{3}U_{\alpha \beta
}^{\pm 1}F\left( R_{\pm },a,U\right) dadU
\]%
which has the clear property $H_{\alpha \beta }^{-}=H_{\beta \alpha }^{+}$ ($%
U_{\alpha \beta }^{-1}=U_{\beta \alpha }$)\qquad

As it was pointed out above this part of the bias for wormholes resembles
formally that for mirrors. To see this analogy we evaluate now the bias for
a gas of mirrors. This case can be formally obtained by setting $R_{+}$ $=$ $%
R_{-}$ $=$ $R$ and $U_{\alpha \beta }$ $=$ $\delta _{\alpha \beta }$. Then
from (\ref{p}) and (\ref{H}) we get
\begin{equation}
b_{0}\left( r\right) =\nabla _{\alpha }\left( h\left( \vec{r}\right) \frac{%
r^{\alpha }}{r^{3}}\right) =\frac{\partial h\left( \vec{r}\right) }{\partial
r^{\alpha }}\frac{\partial \left( -1/r\right) }{\partial r^{\alpha }}+4\pi
h\left( 0\right) \delta \left( \vec{r}\right)  \label{bm}
\end{equation}%
where $h\left( R\right) =\int a^{3}F\left( R,a\right) da$, $F\left(
R,a\right) $ is the distribution of mirrors analogous to (\ref{F}), and we
used the property $\nabla ^{2}\left( -1/r\right) $ $=$ $4\pi \delta \left(
r\right) $. From (\ref{bm}) we see that the bias $b\left( r\right) $
acquires a non-trivial dependence on the radius $r$ only due to the local
inhomogeneity of the gas (i.e., the first term in (\ref{bm})$\ \sim \partial
h\left( \vec{r}\right) $), while in the case of a homogeneous distribution $%
\overline{F}\left( R,a\right) =nf\left( a\right) $ we find $\overline{h}%
\left( r\right) =n\overline{a^{3}}$ ($n$ is the density of mirrors), the
firs term in (\ref{bm}) disappears and, therefore, the mean bias $\overline{b%
}_{0}\left( r\right) $ reduces merely to the renormalization of the point
source $M^{\prime }/M=\left( 1+4\pi n\overline{a^{3}}\right) $ which
corresponds to the case $\varepsilon $ $=$ $1/\left( 1+4\pi n\overline{a^{3}}%
\right) $ $<$ $1$.

In the case of wormholes the bias $b_{0}\left( r\right) $ has the same
structure. Indeed, from (\ref{H}) we see that $b_{0}\left( r\right) =\nabla
_{\alpha }\left( f^{\alpha }\left( r\right) /r\right) $ with some vector $%
f^{\alpha }\left( r\right) $ defined by the integral in (\ref{H}) and if we
assume isotropic distribution of wormholes this vector can be proportional
to the radius only, i.e., $f^{\alpha }\left( r\right) =r^{\alpha }h\left(
r\right) /r^{2}$, with $h\left( r\right) =(\vec{r},\vec{f}\left( r\right) )$%
. Thus we get the same expression (\ref{bm}) with the function $h\left(
r\right) $ defined from (\ref{H}) by
\[
h\left( r\right) =\int \frac{r^{\alpha }R^{\beta }}{R^{2}}\left[ H_{\alpha
\beta }^{+}\left( \vec{r},\vec{R}\right) +H_{\beta \alpha }^{+}\left( \vec{R}%
,\vec{r}\right) \right] d^{3}R.
\]

We point out that the function $h\left( r\right) $ (together with $F$, $%
H_{\alpha \beta }$) has, in general, quite irregular behavior and require
some averaging out. The smooth halo $\overline{b}_{0}\left( r\right) $
around the point source (i.e., Dark Matter halo e.g., Ref \cite{KT07})
appears due to the local inhomogeneity of the function $\overline{h}\left(
r\right) $ and, therefore, due to the local inhomogeneity of the topological
structure, which is in agreement with (\ref{BT}). We also stress that the
applicability of the expression (\ref{bm}) is restricted by sufficiently
large distances, at which the number of wormholes within the radius $r$ is $%
N(r)$ $=$ $4/3\pi r^{3}n$ $\gg $ $1$. At small distances the density of
wormholes fluctuates strongly, e.g., sufficiently close to a source
wormholes are absent which means that $h\left( r\right) $, $b_{0}\left(
r\right) \rightarrow 0$ and the permeability $\varepsilon $ tends to the
vacuum value $\varepsilon \rightarrow 1$.

\subsection{The case $\protect\varepsilon >1$}

Consider now the second part of the bias (\ref{bk1}). For astrophysical
implications (characteristic scales $L\gg a$) it is sufficient to consider
the approximation $\vec{r}_{\pm 1}\approx \vec{R}_{\pm }$, i.e., throats
(every sphere of the radius $a_{n}$) look like point-like objects and every
ghost image is assumed to be in the center of\ a wormhole. Then, the
topological bias is given by%
\begin{equation}
b_{1}\left( r\right) =\sum\limits_{n}a_{n}\left( \frac{1}{R_{-,n}}-\frac{1}{%
R_{+,n}}\right) \left[ \delta (\vec{r}-\vec{R}_{+,n})-\delta (\vec{r}-\vec{R}%
_{-,n})\right] .  \label{bb}
\end{equation}%
In this approximation the bias does not depend on the matrix $U$ and the
density of wormholes (\ref{F}) reduces to $F\left( R_{\pm },a\right) =\int
FdU$. The homogeneity and isotropy of the topological structure mean that $%
\overline{F}=nF\left( \left\vert \vec{R}_{-}-\vec{R}_{+}\right\vert
,a\right) $ where $n=N/V$ is the number density of wormholes in space. Then
the mean bias can be presented as
\begin{equation}
\overline{b}_{1}\left( r\right) =2n\int \left( \frac{1}{R}-\frac{1}{r}%
\right) f\left( \left\vert \vec{R}-\vec{r}\right\vert \right) d^{3}\vec{R}
\label{bb2}
\end{equation}%
where $f\left( X\right) =\frac{1}{n}\int aF\left( X,a\right) da$ (so that $%
\int f\left( x\right) d^{3}x=\overline{a}$) and which in the Fourier
representation $b\left( k\right) =\left( 2\pi \right) ^{-3/2}\int b\left(
r\right) e^{-ikr}d^{3}r$ takes the simplest form
\begin{equation}
\overline{b}_{1}\left( k\right) =2n\frac{4\pi \left( f\left( k\right)
-f\left( 0\right) \right) }{k^{2}}.  \label{b_k}
\end{equation}%
The topological permeability is then given by $\varepsilon (k)$ $=$ $%
1/\left( 1+b\left( k\right) \right) $ $=$ $1$ $-$ $b\left( k\right) /\left(
1+b\left( k\right) \right) $.

Thus, for a specific distribution of wormholes $f\left( k\right) $ the
relation (\ref{b_k}) defines the mean topological polarizability of space
(the mean bias $\overline{b}\left( k\right) $) in the field of the external
source $\phi _{ext}=-1/r$. We recall that by the construction $d$ is here
defined in the range $d=\left\vert \vec{R}_{+}-\vec{R}_{-}\right\vert \geq
2a $ which means that $f\left( k\right) \rightarrow 0$ as $k>\pi /\overline{a%
}$, while for $k\rightarrow 0$ it gives
\[
\overline{b}\left( k\right) \approx 2n\left( \frac{1}{2}f^{^{\prime \prime
}}\left( 0\right) +...\right) .
\]%
For sufficiently large distances $r\rightarrow \infty $ ($k\rightarrow 0$)
we get $\overline{b}\left( k\right) \approx nf^{^{\prime \prime }}\left(
0\right) $, which defines merely the renormalization of the point source $%
M^{\prime }/M=\left( 1+nf^{^{\prime \prime }}\left( 0\right) \right) $. As
we shall see in what follows, this case corresponds to $f^{^{\prime \prime
}}\left( 0\right) <0$ and, therefore, $\varepsilon =1/\left( 1+nf^{^{\prime
\prime }}\left( 0\right) \right) >1$.

Consider now the simplest example when all wormholes have equal values of $%
d=\left\vert \vec{R}_{-}-\vec{R}_{+}\right\vert =r_{0}$. In this case we can
take $f\left( X\right) $ $=$ $\overline{a}/(4\pi r_{0}^{2})\delta \left(
X-r_{0}\right) $ and find $f\left( k\right) $ $=$ $\overline{a}(2\pi
)^{-3/2}\sin (kr_{0})/(kr_{0})$ which defines the bias in the form
\[
\overline{b}_{1}\left( k\right) =-4n\overline{a}(2\pi )^{-1/2}\frac{1}{k^{2}}%
\left( 1-\frac{\sin \left( kr_{0}\right) }{kr_{0}}\right)
\]%
which for $kr_{0}\ll 1$ gives $\overline{b}_{1}\left( k\right) $ $\approx $ $%
-\frac{4n\overline{a}}{(2\pi )^{1/2}}\frac{1}{6}r_{0}^{2}(1$ $-$ $\frac{1}{20%
}\left( kr_{0}\right) ^{2}$ $+$ $...)$. Thus, we see that such bias produces
a negative halo around a point source with the density

\[
\overline{b}_{1}\left( r\right) =-\frac{n\overline{a}}{rr_{0}}\left(
\left\vert r_{0}-r\right\vert +r_{0}-r\right) =-2n\overline{a}\left( \frac{1%
}{r}-\frac{1}{r_{0}}\right) ,~as\ r<r_{0}.
\]%
For $r<r_{0}$ it defines the scale-dependent renormalization of a source
\[
\frac{\delta M\left( r\right) }{M}=4\pi \int_{0}^{r}\overline{b}\left(
r\right) r^{2}dr=-8\pi n\overline{a}M\left( \frac{r^{2}}{2}-\frac{r^{3}}{%
3r_{0}}\right) ,\
\]%
which for $r>r_{0}$ (where $b_{1}=0$) reduces to the constant negative shift
$\ \delta M_{tot}/M=-\frac{4\pi }{3}n\overline{a}r_{0}^{2}$, i.e., we get $%
\varepsilon $ $=$ $1/\left( 1-\frac{4\pi }{3}n\overline{a}r_{0}^{2}\right) $
$>$ $1$. A more general case we obtain when considering an additional
distribution $P\left( r_{0}\right) $ ($\int P\left( x\right) dx=1$) over the
parameter $r_{0}$ $=$ $\left\vert \vec{R}_{-}-\vec{R}_{+}\right\vert $,
which gives merely $\delta M_{tot}/M=-\frac{4\pi }{3}n\left\langle \overline{%
a}r_{0}^{2}\right\rangle $ (where $\left\langle \overline{a}%
r_{0}^{2}\right\rangle =\int \overline{a}x^{2}P\left( x\right) dx$) and
again we find that $\varepsilon >1$.

We see that basic feature of wormholes is that the space possesses a
specific polarizability of the topological origin. Moreover, such a
polarizability exists in gravity as well. From electrodynamics we know that
the polarizability of a medium leads to the screening (partial or total) of
a source. We note that the screening can be effectively described by means
of adding of the "mass-like" term to the Poisson equation $\Delta
\rightarrow \Delta -m^{2}$ which transforms the Green function to the $G\sim
-e^{-mr}/r$. By other words virtual photons or gravitons acquire in such a
medium an effective mass. In particular, in Ref. \cite{sand} it was claimed
that adding of the massive term allows to explain the rotation curve (i.e.,
the amount of dark matter) in any particular galaxy. However to explain the
presence of DM in all galaxies (i.e., the variety of DM halos) the effective
graviton mass should vary in space, ( $m\rightarrow m\left( x\right) $)
which is rather difficult to incorporate in the theory on the very
fundamental level. In the presence of wormholes the bias (\ref{b_k}) can
also be interpreted as such an effective mass-like term which however turns
out to be scale-dependent $m^{2}\left( k\right) =-k^{2}\overline{b}\left(
k\right) /\left( 1+\overline{b}\left( k\right) \right) $. Moreover, the sign
of this term depends essentially on the interplay of the two parts $%
b=b_{0}+b_{1}$, where $b_{1}<0$, while $b_{0}$ may in general have both
signs. Thus, in a particular range of scales $\Delta k$ the effective
mass-like term may have both signs. General consideration (\ref{BT}) shows
that the sign of $b$ depends essentially on the behavior of the physical
volume $V_{phys}\left( r\right) $ (i.e., of the physically admissible region
of space). In the simplistic model considered $V_{phys}\left( r\right)
<4/3\pi r^{3}$ and therefore on sufficiently large distances $b$ is always
positive. We also stress that in the case of a non-trivial topological
structure on the very fundamental level gravitons remain massless (i.e., the
theory does not change at all), while the effective graviton mass appears
merely as the result of the topological polarization effects (i.e., due to
the presence of a gas of wormholes).

\section{Conclusions}

In conclusion we briefly repeat basic results. First of all we have
explicitly demonstrated that a static gas of wormholes leads indeed to the
topological bias of point-like sources which can equally be interpreted as
the presence of a "dark matter" halo around any point source. However in
general, the halo density admits both (positive and negative) signs
depending on scales and the specific features of the distribution of
wormholes. By analogy with the magnetic media we can speak of dia- and para-
susceptibilities of space.

The general geometric consideration has revealed that the sign of the bias
(and that of the halo density) depends on the discrepancy between the
behavior of the volume $V_{phys}\left( r\right) $ of the physically
admissible region of space and that of the coordinate space $V_{coor}\left(
r\right) $ (\ref{BT}) (which was confirmed by the subsequent rigorous
calculations (\ref{bm})). Moreover, a nontrivial halo (the dependence on the
radius $r$) appears only due to the local inhomogeneity of the topological
structure (e.g., see (\ref{bm})). In particular, if we approximate $%
V_{phys}\left( r\right) \sim r^{D}$, then (\ref{BT}) defines the behavior of
the bias as $\overline{b}\left( r\right) $ $\sim $ $\left( 3-D\right)
1/r^{D} $, while at scales where the topological structure crosses over to
homogeneity $V_{phys}\left( r\right) $ $\rightarrow r^{3}$ we get $\overline{%
b}\left( r\right) \rightarrow b\delta \left( r\right) $, i.e., the bias
renormalizes merely the value of the source. We recall the observations
evidence for the value $D\simeq 2$ starting from a few \thinspace $Kpc$ up
to at least $\ 100Mpc$ (e.g., see discussions in Refs. \cite{K06,KT06}).

We note that in the simplistic model considered $V_{phys}\left( r\right)
<4/3\pi r^{3}$ and the total bias has always the positive sign. However,
geometrically one can imagine a more complex topology (e.g., in
multidimensional theories) for which we will get an excess of volume $%
V_{phys}\left( r\right) >4/3\pi r^{3}$ which will lead to a negative bias
and a negative density of Dark halos. It is tempting to relate such a case
to the Dark Energy phenomenon. However, it is clear that this cannot
describe the total fraction of DE. An essential fraction should be also
given by the vacuum zero-point fluctuations\footnote{%
We note that for macroscopic wormholes the Casimir energy density gives only
a tiny contribution to DE.}. Indeed, let us prescribe some finite energy
density to such fluctuations $\sigma _{0}$ (lambda term). In the flat space
the vacuum density should disappear (the exact mechanism of the compensation
or renormalization is not important here), while in the case of a nontrivial
topology some portion of the volume cuts $V_{coor}$ $\rightarrow V_{phys}$
and this gives an additional shift of the vacuum energy density $\sigma _{0}$
$\rightarrow $ $\sigma $ $=$ $\sigma _{0}\left( V_{phys}-V_{coor}\right)
/V_{coor}$, where the sign of $\sigma $ depends on the difference ($%
V_{phys}-V_{coor})$.

\section{Acknowledgment}

For A.A. K this research was supported in part by the joint Russian-Israeli
grant 06-01-72023.

\end{document}